\documentclass[aps,prx,reprint,superscriptaddress,amsmath,amssymb,floatfix]{revtex4-2}

\usepackage{graphicx}
\usepackage{dcolumn}
\usepackage{bm}

\usepackage{amssymb}
\usepackage{hyperref}
\usepackage{cancel}
\usepackage{ulem}

\usepackage{miller}
\usepackage{siunitx}
\DeclareSIUnit \angstrom {\AA}
\DeclareSIUnit \bar {bar}
\usepackage{xcolor}

\begin{document}


\title{A topological material in the III--V family: heteroepitaxial InBi on InAs}

\author{Laurent Nicola\"i}
\email{lnicolai@ntc.zcu.cz}
\affiliation{New Technologies Research Centre, University of West Bohemia, Univerzitni~8, 306~14~Pilsen, Czech Republic}
\author{J\'{a}n Min\'{a}r}
\email{jminar@ntc.zcu.cz}
\affiliation{New Technologies Research Centre, University of West Bohemia, Univerzitni~8, 306~14~Pilsen, Czech Republic}
\author{Maria Christine Richter}
\affiliation{Laboratoire de Physique des Mat\'{e}riaux et des Surfaces, CY~Cergy-Paris Universit\'{e}, LIDYL, CEA, CNRS, 91191 Gif-sur-Yvette, France}
\affiliation{LIDYL, Universit\'{e} Paris-Saclay, CEA, CNRS, 91191 Gif-sur-Yvette, France}
\author{Olivier Heckmann}
\affiliation{Laboratoire de Physique des Mat\'{e}riaux et des Surfaces, CY~Cergy-Paris Universit\'{e}, LIDYL, CEA, CNRS, 91191 Gif-sur-Yvette, France}
\affiliation{LIDYL, Universit\'{e} Paris-Saclay, CEA, CNRS, 91191 Gif-sur-Yvette, France}
\author{Jean-Michel Mariot}
\affiliation{Laboratoire de Chimie Physique--Mati\`{e}re et Rayonnement, Sorbonne Universit\'{e}, CNRS, 4~place Jussieu, 75252 Paris Cedex~05, France}
\author{Uros Djukic}
\affiliation{Laboratoire de Physique des Mat\'{e}riaux et des Surfaces, CY~Cergy-Paris Universit\'{e}, LIDYL, CEA, CNRS, 91191 Gif-sur-Yvette, France}
\author{Johan Adell}
\affiliation{MAX~IV Laboratory, Lund University, Fotongatan~2, 224~84~Lund, Sweden}
\author{Mats Leandersson}
\affiliation{MAX~IV Laboratory, Lund University, Fotongatan~2, 224~84~Lund, Sweden}
\author{Janusz Sadowski}
\affiliation{Institute of Physics, Polish Academy of Sciences, al.~Lotnik\'{o}w~32/46, 02-668~Warsaw, Poland}
\author{J\"{u}rgen Braun}
\affiliation{Department Chemie, Ludwig-Maximilians-Universit\"{a}t M\"{u}nchen, Butenandtstra{\ss}e~11, 81377~M\"{u}nchen, Germany}
\author{Hubert Ebert}
\affiliation{Department Chemie, Ludwig-Maximilians-Universit\"{a}t M\"{u}nchen, Butenandtstra{\ss}e~11, 81377~M\"{u}nchen, Germany}
\author{Jonathan D. Denlinger}
\affiliation{Advanced Light Source, Lawrence Berkeley National Laboratory, 1~Cyclotron Road, Berkeley, CA 94720-8229, USA}
\author{Ivana Vobornik}
\affiliation{TASC Laboratory, Istituto Officina dei Materiali, CNR, Area Science Park--Basovizza, Strada Statale~14, km~163.5, 34149~Trieste, Italy}
\author{Jun Fujii}
\affiliation{TASC Laboratory, Istituto Officina dei Materiali, CNR, Area Science Park--Basovizza, Strada Statale~14, km~163.5, 34149~Trieste, Italy}
\author{Pavol \v{S}utta}
\affiliation{New Technologies Research Centre, University of West Bohemia, Univerzitni~8, 306~14~Pilsen, Czech Republic}
\author{Gavin R. Bell}
\affiliation{Department of Physics, University of Warwick, CV4 7AL Coventry, United Kingdom}
\author{Martin Gmitra}
\affiliation{Institute of Physics, Pavol Jozef \v{S}af\'{a}rik University in Ko\v{s}ice, Park Angelinum~9, 040~01~Ko\v{s}ice,
Slovak Republic}
\affiliation{Institute of Experimental Physics, Slovak Academy of Sciences, Watsonova~47, 040~01~Ko\v{s}ice,
Slovak Republic}
\author{Karol Hricovini}
\email{karol.hricovini@cyu.fr}
\affiliation{Laboratoire de Physique des Mat\'{e}riaux et des Surfaces, CY~Cergy-Paris Universit\'{e}, LIDYL, CEA, CNRS, 91191 Gif-sur-Yvette, France}
\affiliation{LIDYL, Universit\'{e} Paris-Saclay, CEA, CNRS, 91191 Gif-sur-Yvette, France}

\date{\today}

\begin{abstract}
InBi\hkl(001) is formed epitaxially on InAs\hkl(111)-A by depositing Bi on to an In-rich surface. Angle-resolved photoemission measurements reveal topological electronic surface states, close to the $\overline{\rm M}$ high symmetry point. This demonstrates a heteroepitaxial system entirely in the III--V family with topological electronic properties. InBi shows coexistence of Bi and In surface terminations, in contradiction with other III--V materials. For the Bi termination, the study gives a consistent physical picture of the topological surface electronic structure of InBi\hkl(001) terminated by a Bi bilayer rather than a surface formed by splitting to a Bi monolayer termination. Theoretical calculations based on relativistic density functional theory and the one-step model of photoemission clarify the relationship between the InBi\hkl(001) surface termination and the topological surface states, supporting a predominant role of the Bi bilayer termination. Furthermore, a tight-binding model based on this Bi bilayer termination with only Bi--Bi hopping terms, and no Bi--In interaction, gives a deeper insight into the spin texture.
\end{abstract}

\maketitle


\section{Introduction}
\label{sec:intro}
The manipulation and measurement of the electron spin polarization is of central interest for spintronics with a profound impact on nanoelectronics, data storage and computer architectures. Spin-orbit coupling (SOC) plays a key role as it allows to control the spin of charged carriers associated to their orbital motion. Topological materials which have an insulating time-reversal invariant band structure are attractive due to unique properties such as lack of backscattering and spin-momentum locking of surface states~\cite{Hasan2010, Qi2011, Lv2021, Zhang2019, Vergniory2019}. Such properties open new routes for the development of post-complementary metal oxide semiconductor devices and information technology. But, up to now, taking for instance the case of field-effect transistors, the  performances of transistors based on topological insulators (TIs) have not been improved sufficiently to make them competitive with current technologies~\cite{Gilbert2021}. Hence the search for new topological materials compatible with established semiconductor technologies continues.

Since the discovery that Bi$_{1-x}$Sb$_x$ is a three-dimensional (3D) TI~\cite{Hsieh2008}, Bi compounds have inspired strong research efforts which led to the discovery of a lineage of 3D $\mathbb{Z}_2$ TIs~\cite{Lv2021, Hasan2015}. Recently, a particular effort was devoted to search and characterize topological semimetal phases~\cite{Lv2021}. Semimetals are characterized by a very small overlap between the bottom of the conduction bad and the top of the valence band or by a vanishing electronic density of states at the Fermi energy ($E_{\mathrm{F}}$). Semimetals with band crossing / touching in the 3D Brillouin zone (BZ) are generally topologically distinct from other semimetals and are referred to as topological semimetals (TSMs). The TSMs are classified into three groups according to the dimensionality of the band crossings in the momentum space: zero-dimensional (0D) band crossing (encountered in Dirac and Weyl semimetals); twofold / fourfold one-dimensional (1D) band crossing along momentum space lines (topological nodal-line semimetals); and band crossing that is preserved in a 2D surface (nodal-surface semimetals).

III--V materials are a backbone of the semiconductor industry. Hybrid III--V semiconductor / semimetal device structures could exploit topologically protected electronic states with great design flexibility using established material systems. InBi is compatible with standard III--Vs and could pave the way to achieve realistic TI materials for applications~\cite{Fei2019}. 

InBi shares neither the semiconducting character nor the zinc blende / wurtzite structure with the rest of the III--V family. Instead, it is a semimetal having the tetragonal PbO structure (space group $P4/nmm$, n$^\circ$129) ~\cite{Binnie1956, Akgoz1973}. The SOC plays a decisive role in the stabilization of this structure~\cite{Ferhat2006, Zaoui2009}. It is a non-symmorphic material, belonging to the topological nodal-line semimetal group~\cite{Lv2021}, the presence of the nodal line having been demonstrated by angle-resolved photoemission spectroscopy (ARPES) on a cleaved bulk crystal~\cite{Ekahana2017}. Interestingly, in spite of the strong SOC in InBi, the nodal line is preserved which
demonstrates the robust protection by the non-symmorphic symmetry of the crystal structure~\cite{Ekahana2017}. It should also be noted that freestanding InBi\hkl(111) thin films have been predicted to be non-trivial~\cite{Freitas2015, Li2016, Chuang2014} as well as InBi on Si\hkl(111) is supposed to be a 2D TI~\cite{Crisostomo2015, Yao2015}. These predictions, combined with the potential to fabricate semimetal / semiconductor heterostructures within the III--V family, underline that experimental study of InBi thin films is worthy of attention.

Standard techniques have been used to produce InBi single crystals~\cite{Roy1972, Pandya1993, Nishimura2003}. Spontaneous formation of InBi crystals is also observed when Bi is alloyed in III--V semiconductors, such as InAs. Bi substitutes As atomic sites, giving the InAs$_{1-x}$Bi$_x$ alloy system. However, due to a low solubility in the hosting compound, beyond a few \% composition, Bi atoms are no longer incorporated on group V sites and InBi clusters form~\cite{Dominguez2013}. Poor Bi incorporation and very low melting point of InBi $(\approx \SI{110}{\celsius})$, have made the material difficult to grow by conventional molecular beam epitaxy (MBE)~\cite{Keen2014}. We are unaware of any studies of the electronic structure of InBi thin films, which is perhaps a consequence of the challenging growth. 

Here we present a combined experiment / theory study of InBi which shows topological surface states supporting a non-Rashba--type spin texture. InBi epitaxial thin films are formed by depositing Bi on In-rich InAs\hkl(111)-A, producing InBi\hkl(001) domains in conjunction with Bi\hkl(111). Low-energy electron diffraction (LEED), scanning tunnelling microscopy (STM) and X-ray diffraction (XRD) characterize the epitaxy and surface structure. ARPES reveals that InBi\hkl(001) surfaces prepared in this way, unlike the cleavage procedure, host new non-trivial metallic surface states in the vicinity of the $\overline{\rm M}$ high symmetry point of the surface~BZ. Our findings do not agree with the usual trilayer representation, pointing rather to a mixture of two co-existing terminations, Bi bilayers and In monolayers, on a single sample. The experimental data are interpreted with the help of two theoretical approaches. Firstly, \textit{ab initio} calculations are performed within the framework of the density functional theory utilizing the fully relativistic multiple scattering Korringa--Kohn--Rostoker (KKR) Green's function formalism. Secondly, a tight-binding model is developed, which explains the spin texture of the Bi bilayer surface in agreement with experimental and computational results.

The paper is organized as follows. First, in Sec.~\ref{sec:methods}, we present the experimental details, i.e., the sample preparation method and the ARPES/STM measurement conditions, and finally the general theoretical approach. In Sec.~\ref{sec:results}, the experimental results are given and discussed in the light of our modelling calculations. Our findings and their impact are summarized in Sec.~\ref{sec:conclusion}.

\section{Experimental and theoretical methods}
\label{sec:methods}
\subsection{Sample preparation}
\label{subsec:sample}
The substrates used all come from an InAs\hkl(111) \textit{n}-type wafer (Wafer Technology Ltd., UK), doped by sulphur with a carrier concentration of \SI{3e18}{\per\cubic\centi\metre}, polished on both sides.  The substrates were first outgassed in an ultra-high vacuum (UHV) ($\approx \SI{2e-10}{\milli\bar}$); they then underwent 2 to 3 cycles of successive ion bombardment (Ar$^+$, \SI{600}{\electronvolt}) and annealing at $\approx \SI{400}{\celsius}$ (monitored by infrared pyrometer with the emissivity set to $\epsilon=0.33$), each stage having a duration of 20 to~\SI{30}{\min}. The InAs\hkl(111) surface can be cation (In) terminated, namely the A side, or anion (As) terminated, referred to as the B side, due to the lack of inversion symmetry of the InAs crystal (see e.g., Ref.~\cite{Yu2003}). We here present results obtained with only InAs-A type substrates. The LEED patterns of the clean InAs\hkl(111)-A surface show a 2$\times$2 reconstruction, in agreement with previously reported investigations~\cite{Andersson1996}.
An ultrathin film of approximately 20 bilayers of Bi was slowly evaporated onto the substrates with a growth rate of 1~monolayer in about 4 minutes using a Knudsen cell. The low deposition rate of Bi and the presence of an In excess at the on sputter-annealed InAs correspond to ``In-rich'' growth conditions by MBE, as described by Keen \textit{et al.}~\cite{Keen2014}.

\subsection{ARPES, STM and XRD measurements}
\label{subsec:ARPES_STM}
This sample preparation procedure was used numerous times in experiments performed at the APE-LE beamline of the Elettra synchrotron radiation facility (Trieste, Italy) and at the MERLIN beamline of the Advanced Light Source facility (Berkeley, CA, USA), where the measured electronic band structure was similar. The ARPES spectra were acquired at the APE-LE beamline with a Scienta SES~2002~\cite{Panaccione2009} and at the MERLIN beamline with a Scienta R4000 analyzer~\cite{Reininger2007}. 
The STM chamber (descrived in ref.~\cite{Panaccione2009}) which is directly connected through UHV to the ARPES chamber at the APE-LE beamline, allowed in-situ
real space characterizations at room temperature for the same Bi/InAs system as measured by ARPES, providing for the filtered~\cite{Horcas2007} images shown in Figs.~\ref{fig:Fig1},\ref{fig:Fig2}.
The  ARPES measurements shown in Figs.~\ref{fig:Fig3}(a) and \ref{fig:Fig3}(c) were performed at photon energies of \SI{87}{\electronvolt} and \SI{20}{\electronvolt}, respectively. During the measurements, the pressure did not exceed~\SI{2e-10}{\milli\bar}.
The XRD patterns of the Bismuth film deposited on a single-crystalline InAs substrate is shown in Fig.\ref{fig:Fig1}(f). An automatic powder X-ray diffractometer X’Pert Pro equipped with an ultra-fast linear semiconductor detector PIXcel was used. Copper K radiation ($\lambda$ = 0.154 nm) was used as an X-rays source. XRD patterns were recorded from 15 to 75 degrees in 2$\theta$ scales. A symmetric (Bragg-Brentano) $\theta-\theta$ geometry, where the X-rays angle of incidence to the sample surface is equal to the angle of reflectance, was used. In this configuration the lattice planes (hkl) where X-ray diffraction takes place are parallel to the sample surface. This condition was kept the same during the whole measurement.To find out the real structure of the film, line profile analysis was used. The procedure, proposed by Langford \cite{Langford1978}, based on a Voigt function analysing the breadths of diffraction lines (FWHM and integral breadth $\beta$=area/$I_0$ where $I_0$ is the intensity at the line profile maximum) was used. Alumina powder from NIST was used as an instrumental standard. To perform the qualitative phase analysis of the films, XRD standards from ICCD database (International Center for Diffraction Data) were used. 

\subsection{Modelling framework}
\label{subsec:Mod}
On the theoretical side, electronic structure \textit{ab initio} calculations were performed using the SPR-KKR package~\cite{Ebert2011}. It is based on the Korringa--Kohn--Rostoker method which uses the Green's function formalism within the multiple scattering theory. The package is based on the fully relativistic Dirac equation, thus fundamentally taking into account the relativistic effects, such as SOC.
The density functional theory (DFT) calculations were done using the local density approximation for both Bi and InBi. The ARPES spectra for the semi-infinite Bi\hkl(111) crystal were calculated within the one-step model of photoemission~\cite{Braun2018}. The energy of the photon set for the calculation presented here was \SI{22}{\electronvolt}. Calculations for Bi and InBi were done in the semi-infinite crystal framework, with a 2D thin film surface created on top in the case of InBi, using the screened KKR method.

To study in detail the SOC effect on the spin textures of the in-gap surface states near the time-reversal invariant momenta $\overline{\rm M}$ (see below), we propose a tight-binding model for the surface Bi bilayer. For this purpose, only electron spin flip hopping terms among the Bi atoms are considered: between nearest neighbors $\lambda_{\rm nn}$, inter-layers, and next nearest neighbors $\lambda_{\rm nnn}$, intra-layer, as sketched in Fig.~\ref{fig:Fig5}(b). The influence of all underlying layers, including the In, is modelled with an on-site potential $V$ acting on the interfacial Bi layer. The derived Hamiltonian in the vicinity of $\overline{\rm M}$ is written as:
\begin{equation}\label{equation}
	{\cal H}_{\rm so} = ({\bf\Omega}_0 + {\bf\Omega}_1 + {\bf\Omega}_3)\cdot \mathbf{\hat{s}},
\end{equation} 
where the $\Omega$ terms are the $k$-dependent SOC fields and $\hat{\bf{s}}$ is the vector of the Pauli matrices. Calculation of the spin expectation values based on the model Hamiltonian ${\cal H}_{\rm so}$, shown in Fig.~\ref{fig:Fig5}(c), reproduces the spin vector fields derived from the \textit{ab initio} calculations [see Fig.~\ref{fig:Fig5}(a)]. The first term, ${\bf\Omega}_0$, including only intra-layer interactions, gives a pure Rashba type of spin pattern. However, this term alone is insufficient to describe the overall spin expectation vector field for both rings. We therefore consider two fields ${\bf\Omega}_1$ and ${\bf\Omega}_3$ that include both intra- and inter-layers interactions renormalized by the on-site potential~$V$. This results in a Rashba--Dresselhaus type texture with 4-fold rotational symmetry. More details on the model can be found in the Appendix.

\section{Results and discussion}
\label{sec:results}
\subsection{Epitaxy and surface structure of InBi}
\label{subsec:epitaxy}

\begin{figure*}
	\includegraphics[width=\linewidth]{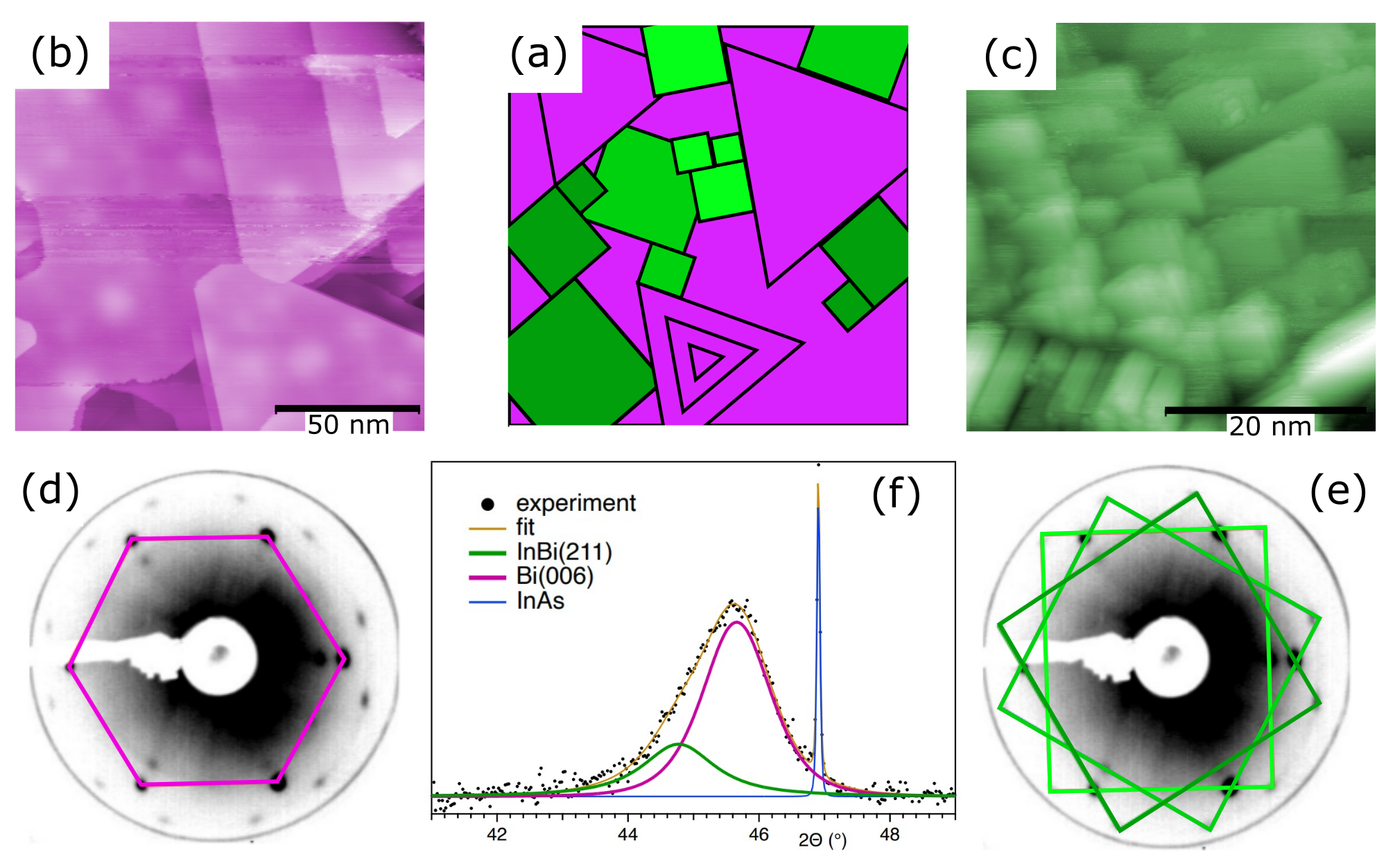}
	\caption{Sample structure. A schematic (a) of the sample highlights three rotational domains of epitaxial tetragonal InBi\hkl(001) [in 3 shades of green] co-existing with epitaxial Bi\hkl(111) [in purple]. STM topographs of (b) Bi and (c) InBi domains show step-terrace structures of the expected symmetries.  LEED patterns at \SI{33}{\electronvolt} highlight the coexisting Bi\hkl(111) and InBi\hkl(001) domains in (d) and (e) respectively. The XRD pattern (f) is shown with fitted components due to the InAs substrate [blue], Bi\hkl(006) [purple] and InBi\hkl(211) [green].
}
\label{fig:Fig1}
\end{figure*}

Samples were prepared by depositing approximately 20 bilayers of Bi on to In-terminated InAs\hkl(111)-A in UHV. 
The appearance of InBi crystals is not very sensitive to the thickness of the deposited Bi layer that can vary from about 10 to several tens of bilayers.
Details are given in Sec.~\ref{sec:methods}. 
The film structure obtained is summarized schematically in Fig.~\ref{fig:Fig1}(a). A Bi\hkl(111) epilayer (purple triangles) exist alongside InBi\hkl(001) domains in three epitaxial orientations (green squares). The InBi domains are mutually separated by 30$^\circ$, which arises from the threefold symmetry of the InAs\hkl(111) substrate. The LEED patterns [Figs.~\ref{fig:Fig1}(d) and \ref{fig:Fig1}(e)], STM images [Figs.~\ref{fig:Fig1}(b) and \ref{fig:Fig1}(c)] and XRD pattern [Fig.~\ref{fig:Fig1}(f)] are all consistent with such a structure. Bi\hkl(111) regions show characteristic triangular step-terrace structures [Fig.~\ref{fig:Fig1}(b)], while InBi\hkl(001) domains appear with perpendicular step-terrace arrays [Fig.~\ref{fig:Fig1}(c)]. In LEED, the hexagonal pattern of Bi\hkl(111) is highlighted in Fig.~\ref{fig:Fig1}(d), while the InBi\hkl(001) domains appear as three squares rotated by 30$^\circ$ as highlighted in Fig.~\ref{fig:Fig1}(e). Both parts of the LEED pattern have in-plane lattice spacing consistent with these materials, and no fractional order spots are observed. 
Out-of-plane lattice parameters obtained by XRD are also consistent with the presence of epitaxial Bi and InBi crystals, as shown in Fig.~\ref{fig:Fig1}(f).
XRD was intended to prove the presence of InBi crystals. Their orientation is evidenced from several other sources and confirmed by surface sensitivity techniques, as discussed below.

\begin{figure*}
	\includegraphics[width=\linewidth]{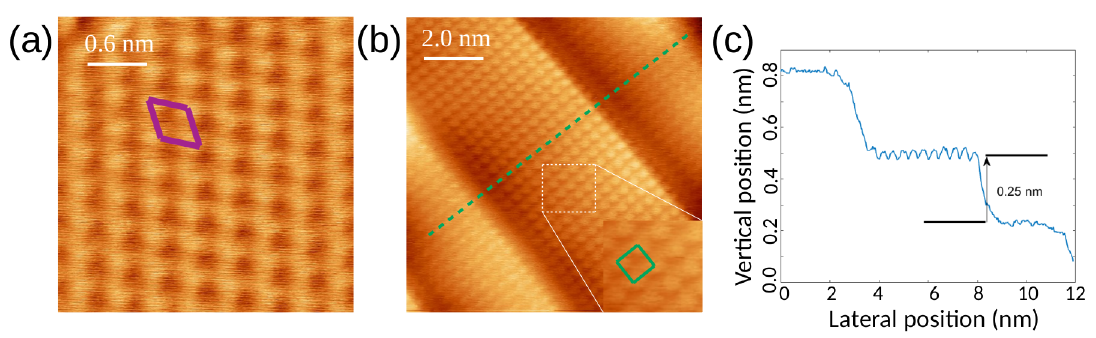}
	\caption{STM images of (a) Bi\hkl(111) and (b) InBi\hkl(001), the latter including a magnified section. The purple parallelogram and green square highlight the respective surface symmetries. A height profile across the InBi step-terrace structure [green dotted line in~(b)] is shown in~(c).
}
\label{fig:Fig2}
\end{figure*}

Higher resolution STM images are shown in Fig.~\ref{fig:Fig2}. Panel~(a) highlights the expected hexagonal pattern for Bi\hkl(111), while the square unit mesh of InBi\hkl(001) is clearly visible in panel~(b). The unit meshes are identical to the bulk--terminated structure, i.e., no surface reconstruction is present, in agreement with LEED. A height profile across the step-terrace structure of InBi\hkl(001) is shown in panel~(c); it clearly highlights a step height of around \SI{0.25}{\nano\metre}. This is half the InBi unit cell height ($c/2$) and implies that the adjacent terraces must be terminated with In or Bi atoms. The structural nonequivalence of adjacent terraces is supported by the stronger or weaker atomic-scale corrugations clearly visible in both the height profile and image. Although the InBi regions have high step density (Figure~\ref{fig:Fig1}c) it was possible to clearly measure single steps in images from different areas. A histogram of step heights is given in the Supplemental Material, Figure S2, which confirms the dominance of half-unit-cell height steps on these samples. We discuss the implications for electronic structure in Sec.~\ref{subsec:GSA}, but here point out that such behavior is unique among III--V materials. While III- and V-terminated surface domains can co-exist in narrow temperature windows during surface reconstruction transitions~\cite{Yamaguchi1995}, III--V semiconductors show a single termination under normal MBE growth conditions. A lack of surface reconstruction is also unique to the \hkl(001) surface of InBi among the III--V family.

The growth of InBi during deposition of pure Bi requires a source of In at the surface. The InAs\hkl(111) surface is polar, with the \hkl(111)-A surface being In-terminated and \hkl(111)-B being As-terminated. Furthermore, Ar$^+$ ion sputter-annealed InAs surfaces readily become enriched with In owing to both preferential sputtering of As and the higher thermal desorption rate of As compared to In~\cite{Bell1996}. In the present \hkl(111)-A case, excess In from the surface cleaning process provides a reservoir of material which can react with incoming Bi to form InBi (see Supplemental Material, Figure S4). 
Note that we have also evaporated Bi on a InAs\hkl(111)-B substrate without observing InBi formation. This highlights the crucial role of the In-rich surface environment.

The growth of InBi with \hkl[001] surface orientation, corresponding to the natural bulk cleavage plane, presumably minimizes the interfacial and surface energies. The growth of InBi nanocrystals by MBE has also been studied when doping InAs with Bi beyond the solubility limit~\cite{Dominguez2013}. It was observed that co-deposition of In, Bi, and As led to InBi nanocrystals embedded in an InAs$_{0.95}$Bi$_{0.05}$ matrix. Despite the fact that the substrate was InAs\hkl(001), the nanocrystals were not oriented with InBi\hkl[001] parallel to InAs[001]. Instead, the InBi\hkl[001] direction was aligned with InAs\hkl[111], as shown by Dominguez et \textit{al.} \cite{Dominguez2013} with transmission electron microscopy, the same epitaxial orientation that we observe in the present study. The nanocrystals were strained (+6.6\% for~$a$; +9.6\% for~$c$), becoming closer to cubic symmetry but remaining tetragonal. It was argued that this orientation minimized the strain energy with respect to the matrix, when compared to a pseudo-cubic InBi polymorph with $c$ aligned along the InAs\hkl[001] growth axis~\cite{Dominguez2013}. Our results support the idea that matching the InAs\hkl(111) plane to the InBi\hkl(001) plane is a general phenomenon. In our case, XRD indicates a closer match between the InBi films and the bulk crystal ($-0.3$\% for~$a$; +1.15\% for~$c$) and no distortion towards a cubic symmetry, as expected for free films compared to embedded nanocrystals.

\subsection{ARPES measurements of Bi and InBi}
\label{subsec:ARPES}

\begin{figure*}
	\centering
	\includegraphics[width=\linewidth]{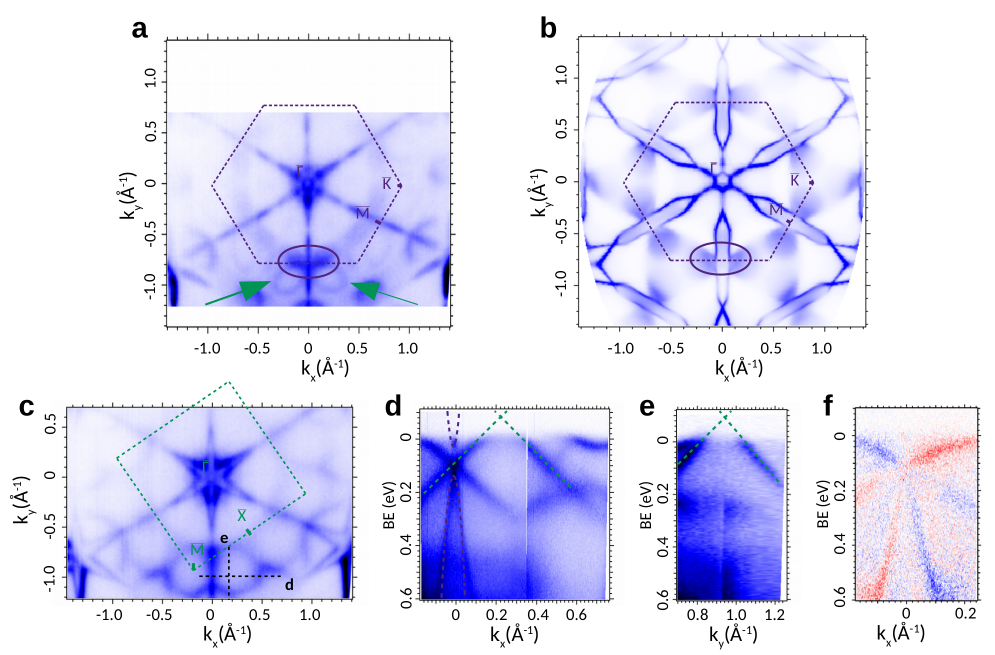}
	\caption{Experimental and theoretical band dispersions for Bi\hkl(111)/InAs\hkl(111)-A. 
	(a) Fermi surface at $h\nu=\SI{87}{\electronvolt}$ of a Bi\hkl(111)/InAs\hkl(111)-A system after depositing $\approx 20$ bilayers of Bi on the InAs substrate. The purple dotted lines delimit the Bi\hkl(111) surface~BZ. States discussed in the text are highlighted by green arrows [InBi] and purple ovals [Bi, also in~(b)]. 
	(b) Calculated Fermi surface at $h\nu=\SI{20}{\electronvolt}$ using the one-step model of photoemission for Bi crystal. 
	(c) Constant energy surface at \SI{100}{\milli\electronvolt} binding energy. The green dotted lines delimit the InBi\hkl(100) surface~BZ.
	(d,e) Dispersions measured respectively along the black horizontal and vertical dotted lines in~(c). The green dotted lines sketch the prolongation of InBi surface states suggesting a Dirac point above~$E_{\mathrm{F}}$. Purple dotted lines highlight the dispersion of Bi valence and conduction bands in the vicinity.
	(f) Circular dichroism spectrum measured along the $k_{x}$ direction, with $k_{y} \approx \SI{1.08}{\per\angstrom}$.}
	\label{fig:Fig3}
\end{figure*}

ARPES constant energy mapping at $E_{\mathrm{F}}$ is displayed in Fig.~\ref{fig:Fig3}(a). The main features of the ARPES spectrum exhibit a hexagonal symmetry, sketched with the purple dotted lines. One can see the already well-described Rashba-like spin-orbit split surface states corresponding to the Bi\hkl(111) crystal~\cite{Hofmann2006, Ast2003, Ohtsubo2012, Koroteev2004}, in particular along the $\overline{\rm \Gamma}$--$\overline{\rm M}$ direction. Our KKR calculations performed for a Bi\hkl(111) monocrystal reproduce the overall hexagonal symmetry with these split states [Fig.~\ref{fig:Fig3}(b)]. Closer inspection of the measured Fermi surface reveals however new features for this particular monocrystal prepared InAs\hkl(111)-A as compared to previous reports on Bi\hkl(111) bulk crystals: close to the $\overline{\rm M}$ point, one observes new states shaped in a butterfly-like pattern [marked with a purple loop in Fig.~\ref{fig:Fig3}(a)]. We suggest that these electronic states, of bulk nature, are due to the filling of the bottom of the Bi conduction band with electrons  supplied by the high density electron gas of the accumulation layer present on the pristine InAs\hkl(111) substrate~\cite{King2010}. Our calculation reproduces clearly these states when electron doping is taken into account [Fig.~\ref{fig:Fig3}(b)].

In the experimental Fermi surface mapping [Fig.~\ref{fig:Fig3}(a)] additional electronic bands can be found further beyond the $\overline{\rm M}$ point of the Bi\hkl(111) crystal. These bands, exhibiting ring-like shapes, are shown by arrows and are referred as ``ring states'' in the following text. In Fig.~\ref{fig:Fig3}(c), a constant energy surface at \SI{100}{\milli\electronvolt} below $E_{\mathrm{F}}$ shows a closer view of these ring states that are of surface type, as no shift is observed when changing the photon energy (see Supplemental Material, Figure S3). The band dispersion along the dotted black lines in Fig.~\ref{fig:Fig3}(c) is linear and is converging towards a Dirac-like point at $\approx\SI{100}{\milli\electronvolt}$ above $E_{\mathrm{F}}$, as displayed in Figs.~\ref{fig:Fig3}(d) and \ref{fig:Fig3}(e). Note that the steep bands dispersing within the $\pm\SI{0.1}{\per\angstrom}$ $k_{\parallel}$ range at the binding energy (BE)of \SI{0.6}{\electronvolt} and crossing close to $E_{\mathrm{F}}$ in Fig.~\ref{fig:Fig3}(d) belong to the Bi\hkl(111) monocrystal~(see Ref.~\cite{Ohtsubo2013}).

A total of 12 ring states is found if the entire BZ is taken into account (see Supplemental Material, Figures S1 and S5).
In the light of the LEED pattern in Fig.~\ref{fig:Fig1}(d) a straightforward interpretation is to place the rings at the corners of the tetragonal BZ of the InBi crystal, referred to as $\overline{\rm M}$ points [see Fig.~\ref{fig:Fig4}(a)], as indicated by the green square in Fig.~\ref{fig:Fig3}(c). Consequently, due to this interpretation the suggested Dirac point is situated exactly at the $\overline{\rm M}$ point of the InBi\hkl(001) crystal. An additional argument in favor of the topological nature of the ring states comes from the circular dichroism spectrum shown in Fig.~\ref{fig:Fig3}(f) which suggests their strong spin polarization. However, at this point one must be careful as it is known that the dichroism cannot be directly connected to the spin nature in a one-to-one model due to the matrix element effects of the photoemission process~\cite{SanchezBarriga2014, Kim2012, Scholz2013, Vidal2013, Xu2015}. Yet, in a previous work, thanks to  theoretical support, we have already shown that dichroic effects can be used as a mean to detect spin polarized trends around the $\overline{\rm \Gamma}$ point of Bi\hkl(111)~\cite{Nicolai2019}. Therefore, in order to obtain a deeper insight into the spin nature of the ring states, the following section is dedicated to a comparison of the experimental observations to the predictions of \textit{ab initio} calculations. 

\subsection{Ground state analyses of InBi}
\label{subsec:GSA}
\begin{figure*}
	\includegraphics[width=\linewidth]{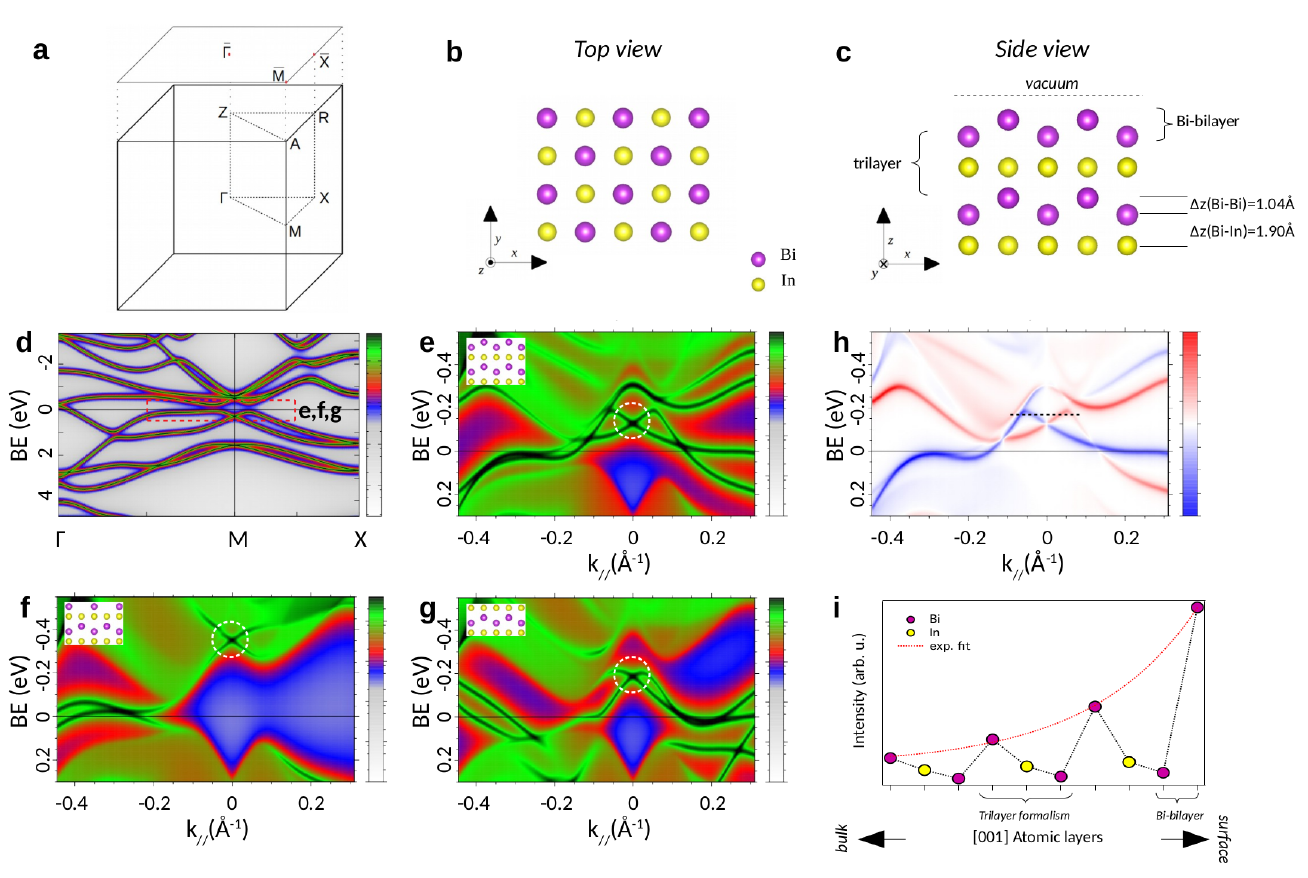}
	\caption{Crystal configuration and calculated electronic structure of the InBi\hkl(001) surface. 
	(a) Surface and bulk BZs of the tetragonal structure. 
	(b, c) Top and side views of the InBi crystal with the bismuth bilayer termination on the \hkl(001) surface. 
	(d) Bloch spectral function along the $\rm \Gamma$--$\rm M$--$\rm X$ directions in bulk InBi. 
	(e, f, and g) Projected Bloch spectral functions along the $\overline{\rm \Gamma}$--$\overline{\rm M}$--$\overline{\rm X}$ directions for the Bi bilayer, trilayer, and In-terminated surface, respectively. For each figure the surface termination is shown in an inset. The \textbf{k} vector and energy spans correspond to the red dotted rectangle in~(d). The surface band crossing in~(e), indicated by the white dotted circle, behaves like a Dirac point. 
	(h) Spin polarization for Bi bilayer terminated surface. 
	(i) Spectral intensity, in the case of Bi bilayer termination, determined at the Dirac point as a function of the atomic plane position from the surface.}
	\label{fig:Fig4}
\end{figure*}

The \hkl(001) surface and bulk BZs of the tetragonal crystal structure are depicted in Fig.~\ref{fig:Fig4}(a) and the top and side views of the InBi crystal with the bismuth bilayer termination on the \hkl(001) surface in Figs.~\ref{fig:Fig4}(b) and \ref{fig:Fig4}(c). In real space the Bi atom is surrounded by 4 In atoms forming a tetrahedron. As the In--Bi distance is the shortest (\SI{3.106}{\angstrom}), compared to the Bi--Bi distance (\SI{3.687}{\angstrom}), InBi is conventionally described as a stacking of Bi--In--Bi trilayers with weak Bi--Bi bonds~\cite{Binnie1956, Ekahana2017} [left brace in Fig.~\ref{fig:Fig4}(c)]. However, from another point of view, the same structure can be seen as a stacking of Bi bilayers (right brace), separated by In monolayers. The STM topography clearly shows terraces separated by \SI{0.25}{\nano\metre}, i.e., $\approx c/2$. Splitting the crystal in the middle of the Bi bilayer can only occur once per unit cell, giving a step height of~$c$ which is inconsistent with STM data. The ``alternate element stacking'' picture is consistent with $c/2$ step heights provided that both In monolayer-terminated and Bi bilayer-terminated terraces can coexist. However, the STM images do not allow us to identify and quantify the total area covered by In-terminated and Bi-terminated terraces across the tetragonal domains. As a consequence, it becomes essential to investigate all possible surface terminations and evaluate their intrinsic impact on the electronic band structure.

When including the surface in the calculation the situation around the gap changes significantly. In Figs.\ref{fig:Fig4}(e), \ref{fig:Fig4}(f), and \ref{fig:Fig4}(g), we show projected Bloch spectral functions along the $\overline{\rm \Gamma}$--$\overline{\rm M}$--$\overline{\rm X}$ directions for the Bi bilayer, trilayer and In-terminated surface, respectively. For more clarity an inset displays the surface termination in each figure. Clearly, the presence of the surface generates a complex network of surface states, appearing as black lines in this logarithmic color scale. These surface states are degenerate at the $\overline{\rm M}$ point due to time-reversal and crystal symmetries. These surface states cross at the $\overline{\rm M}$ point, indicating a potential chiral spin texture.

Depending on the surface termination the surface states occupy the bulk band gap or are situated at the edges of the projected bulk bands. As a remark, we notice that the band gap remains open throughout the $k_z$ direction (see Supplemental Material, Figure S6). A closer look reveals the most prominent Dirac-like point in the middle of the bulk band gap, indicated by the white dotted circle, for the Bi bilayer-terminated surface, in Fig.~\ref{fig:Fig4}(e). This calculated surface state can be related to ARPES measurements shown in Figs.~\ref{fig:Fig3}(d) and \ref{fig:Fig3}(e). As seen here, the surface band crossing moves to higher energies, on the upper edge of the bulk band gap for the other surface terminations [Figs.~\ref{fig:Fig4}(f) and \ref{fig:Fig4}(g)].
Surprisingly, such bands were not observed in recent ARPES experiments~\cite{Ekahana2017} where the clean surface was obtained by cleaving a bulk crystal. We suggest that also in this case the surface cannot be terminated by a Bi monolayer, as it is generally considered that the cleavage plane spans between trilayers. On the contrary, we suggest that there is also a mixing of Bi and In terminations. MBE could allow preparation of InBi\hkl(001) with fully Bi bilayer terminated surface, however, up to date, there is no such report in the literature. Therefore, the deposition on In-rich InAs\hkl(111)-A surface, as described here, appears to be a unique way to grow InBi crystalline films.

The spin-resolved Bloch spectral function in Fig.~\ref{fig:Fig4}(h) shows more in detail the spin texture in the vicinity of the $\overline{\rm M}$ point. The highlighted surface bands display a spin momentum locking, therefore corroborating the circular dichroism effect reported above [Fig.~\ref{fig:Fig3}(f)].

A complementary insight into the electronic properties induced by the presence of the Bi bilayer termination is given by the evolution of the spectral intensity, determined at the Dirac point, as a function of the atom location measured from the surface to the bulk [Fig.~\ref{fig:Fig4}(i)] depicting the contribution of one atom for each atomic plane. The contributon is a result of the integration of the intensity in the vicinity around the Dirac point as observed in Fig.~\ref{fig:Fig4}(e). Clearly, the intensity at the Dirac point is prominently generated by the Bi excess layer, i.e., the upper layer of the Bi bilayer. This intensity propagates resonantly into the bulk and is mainly harboured by the upper Bi atom from each bilayer. We have evaluated an equivalent tendency for the other two surface terminations (see Supplemental Material, Figure S7). Nonetheless, both exhibit a smaller maximum intensity, relatively reduced by a factor of about~2.8. So, the Bi bilayer termination not only induces the surface band crossing approximately in the middle of the bulk band gap at the $\overline{\rm M}$ point, but as well with the most prominent intensity. Therefore we will delve further into the Bi bilayer termination case. 

\subsection{Spin-orbit coupling effect on the in-gap surface states}
\label{subsec:SOC}

\begin{figure*}
	\includegraphics[width=\linewidth]{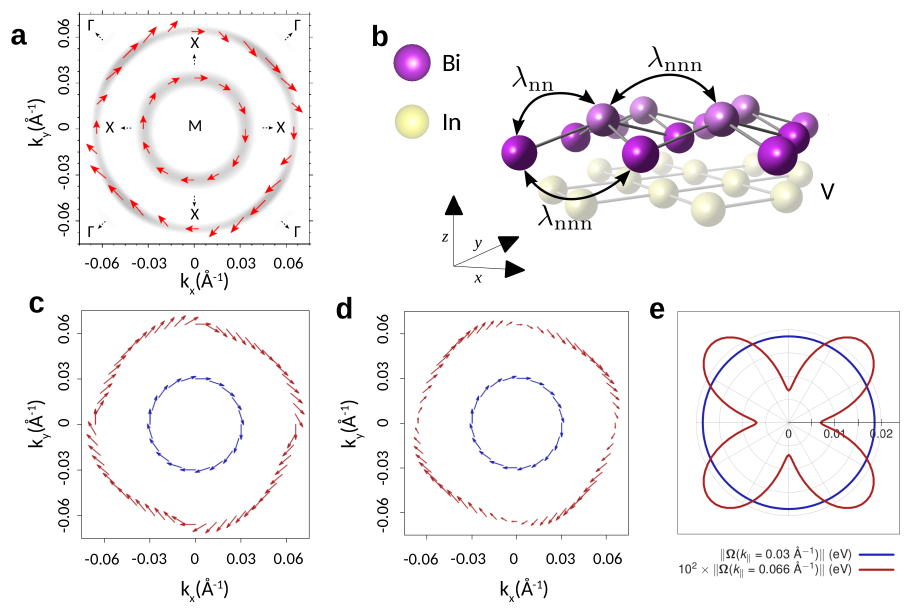}
	\caption{Layer contributions and spin vector fields. (a) SPR-KKR Bloch spectral intensities for isoenergetic plan around the $\overline{\rm M}$ point with spin of bands of interest. (b) Sketch of the atomic model of the Bi bilayer with SOC hopping terms $\lambda_{\rm nn}$ and $\lambda_{\rm nnn}$ between nearest- and next nearest-neighbors and on-site potential $V$ on the interfacial Bi layer. No hopping term is therefore considered for the In atoms. (c) Plot of the spin distribution for both rings around the $\overline{\rm M}$ point obtained from tight-binding model fit for the bands with wave vector $k_{\parallel}=\SI{0.03}{\per\angstrom}$ (blue) and $k_{\parallel}=\SI{0.066}{\per\angstrom}$ (red). (d) Corresponding SOC fields $\bf{\Omega}(\bf{k})$. (e) Polar plots of $\bf{\Omega}(\bf{k})$.}
	\label{fig:Fig5}
\end{figure*}

In this section we will discuss only the Bi bilayer surface termination case. The spin texture of the electron bands, calculated using the SPR-KKR package, is revealed [see Fig.~\ref{fig:Fig5}(a)] in a constant energy mapping, above the band crossing, as indicated by the black dotted line in Fig.~\ref{fig:Fig4}(h). An inner and outer ring electron band with different spin texture can be identified. The inner ring depicts a Rashba-like spin helicity. 
However, the outer ring displays a more complex four-fold symmetric spin pattern with both radial and tangential spin components. Pure tangential spins are only visible along the symmetry axes: the $\overline{\rm \Gamma}$--$\overline{\rm M}$ and $\overline{\rm X}$--$\overline{\rm M}$ directions. For the other directions in momentum space the spin expectation values reveal significant radial contributions.

As mentioned above, our tight-binding model predicts a Rashba--Dresselhaus type texture with 4-fold rotational symmetry [see Fig.~\ref{fig:Fig5}(e)], giving rise to a radial spin expectation component except at the momenta along the main axes pointing towards the $\overline{\rm X}$ and $\overline{\rm \Gamma}$ points [see Fig.~\ref{fig:Fig5}(d)].
The reproduction of the electronic structure with this model, i.e., a well connected Bi bilayer relatively separated from the In layers, further dismantles the trilayer stacking image of InBi. The intra-link between two close Bi atoms was generally perceived as a weak Van der Waals like bond, prone to being cut when cleaving the sample in the \hkl[001] orientation. The electronic structure shows that two neighboring Bi atoms are more linked than first thought. As a result, it is not surprising to identify both In-terminated and Bi-terminated areas on a InBi\hkl(001) surface. As such, we also suggest that cleaved crystal will display both type of surface termination. In other words, InBi would be, to our knowledge, the first identified III--V family member that does not show a mono-atomic composition at its cleaving plane. 

InBi presents similarities with a new stable phase of Bi crystallizing in space group $I4/mmm$ (n$^\circ$139), Bi-139, that has been theorized to be a topological crystalline insulator~\cite{Munoz2016}. 
It has been reported that for Bi-139, around the $\overline{\rm M}$ point, the predicted spin distribution should follow a non pure Rashba behavior~\cite{Munoz2016}, like in our InBi system. Bi-139 has in total 3 mirror planes, including having a mirror plane along \hkl[001], unlike InBi, while sharing the two remaining symmetry planes. This would hint that the In layers intercalated between the Bi bilayer of InBi are preventing a \hkl[001] mirror plane.

\section{Conclusion}
\label{sec:conclusion}

Deposition of Bi on to the In-terminated InAs\hkl(111)-A substrate leads to the formation of InBi\hkl(001) epitaxial domains that coexist with hexagonal Bi\hkl(111), confirmed by LEED, STM, and XRD. STM further shows that InBi\hkl(001) grown in this way has co-existing In- and Bi-terminated terraces, which is unique among III--V materials. ARPES studies reveal new electronic surface states of InBi\hkl(001) exhibiting a spin momentum locking. Layer-resolved electronic band structure \textit{ab initio} calculations using the SPR-KKR package show that these states are strongly dependent on the InBi\hkl(001) crystal termination, and three are considered. (1)~A Bi bilayer termination brings the predominant intensity contribution at the Dirac point of the surface bands. This atomic configuration is to be compared to the predicted Bi-139~\cite{Munoz2016} with Bi bilayers reflecting the same in-plane symmetries. However, in the Bi-139 crystal there is a mirror plane normal to the \hkl(001) direction between two Bi bilayers. In the InBi crystal two adjacent Bi bilayers are separated by an indium atomic layer which, on the contrary, constitutes a glide plane. (2)~For the indium termination the spectral intensity at the Dirac point is about three times lower compared to the Bi bilayer termination. (3)~In contrast, for the termination claimed to occur for a cleaved InBi crystal~\cite{Ekahana2017}, i.e., a Bi monolayer, we find that the topological states merge with the bulk bands. Consequently, our experimental and theoretical studies rule out the ``trilayer stacking'' picture with Bi monolayer termination. The unique InBi growth process reported here allows control of the topological properties of the InBi\hkl(001) face. A similar behavior of spin texture dependence on the crystal termination has been observed in the Pt$_{3}$Te$_{4}$, characterized by weak Van der Waals bonds~\cite{Fujii2021}. Further MBE-based studies of the InBi epitaxial system could allow both optimization of the atomic surface composition and improved integration with conventional III--V materials. For example, very recent experiments have examined the use of MBE and periodic supply epitaxy for the direct growth of InBi(001) on InSb(001) \cite{Tom2024}.

\begin{acknowledgments}
This publication was supported by the project CEDAMNF with reg.\ no.\ CZ.02.1.01/0.0/0.0/15$\_$003/0000358 and within the project QM4ST with reg.\ no.\ CZ.02.01.01/00/22$\_$008/0004572, co-funded by the ERDF as part of the MŠMT (Czech Republic).
J.-M.M.\ acknowledges support from the European Community's Seventh Framework Programme (FP7/2007-2013) under grant agreements~n$^{\circ}$226716 and 312284. M.G.\ acknowledges financial support provided by Slovak Research and Development Agency provided under Contract No.\ APVV-SK-CZ-RD-21-0114 and by the Ministry of Education, Science, Research and Sport of the Slovak Republic provided under Grant No.\ VEGA 1/0105/20 and Slovak Academy of Sciences project IMPULZ IM-2021-42 and project FLAG ERA JTC 2021 2DSOTECH. This research used resources of the Advanced Light Source, which is a DOE Office of Science User Facility under contract no.\ DE-AC02-05CH11231. This work has been partly performed in the framework of the Nanoscience Foundry and Fine Analysis (NFFA-MUR Italy Progetti Internazionali) facility. The authors thank Thiagarajan Balasubramanian and Craig Polley for scientific discussions, helping us in interpreting the presented set of data.
\end{acknowledgments}

\appendix*
\section{Details on the tight-binding model}
\label{sec:Appendix}
In our tight-binding model, we only consider $p_z$ orbitals for the Bi atoms as supported by the KKR calculations. The energy of the orbitals is set to $\epsilon_0$ controlling the position of the Dirac point. The effect of the InBi substrate on the Bi bilayer is modelled as a momentum independent on-site potential~$V$.
By means of the theory of invariants~\cite{Winkler2003} within the $C_{4v}$ point group, we identified in the vicinity of the $\overline{\rm M}$ point two relevant spin-flip SOC hoppings between the nearest Bi atoms $\lambda_{\rm nn}$ and next-nearest neighbors $\lambda_{\rm nnn}$ linear in momentum (see the sketch in the Fig.~\ref{fig:Fig5}(b). Using L\"owdin partitioning~\cite{Lowdin1951} assuming strong coupling of the bottom Bi layer to the InBi substrate, $V\gg \epsilon_0$, the SOC Hamiltonian near the $\overline{\rm M}$ point can be written as: ${\cal H}_{\rm so} = ({\bf\Omega}_0 + {\bf\Omega}_1 + {\bf\Omega}_3)\cdot \hat{\bf{s}}$, where $\hat{\bf{s}} = (\hat{\bf{x}}\sigma_x + \hat{\bf{y}}\sigma_y)^{\rm T}$ is the vector of the spin Pauli matrices, and the spin-orbit fields are ${\bf\Omega}_0 = 2 \lambda_{\rm nnn} \emph{k}a (\hat{\bf{x}}\sin{\varphi}-\hat{\bf{y}}\cos{\varphi})$, ${\bf\Omega}_1 = \Lambda_1 (\emph{k}a)^3\sin(2\varphi)(\hat{\bf{x}}\cos{\varphi}-\hat{\bf{y}}\sin{\varphi})$, and ${\bf\Omega}_3 = \Lambda_3 (ka)^3 (\hat{\bf{x}}\sin{3\varphi}+\hat{\bf{y}}\cos{3\varphi})$, $\hat{\bf{x}}$ and $\hat{\bf{y}}$ being  the unit vectors along the reciprocal lattice vectors. $\Lambda_1 = 4Q t \lambda_{\rm nn} (\epsilon_0 - V)$ and $\Lambda_3 = 8Q\lambda_{\rm nn}\lambda_{\rm nnn}$ are the SOC strengths, where $Q=\left[\left(\epsilon_0 - V\right)^2-4\emph{k}^2\lambda_{\rm nnn}^2\right]^{-1}$ and $t$ is the orbital hopping between the nearest neighbor Bi atoms, $a$ being the lattice constant (\SI{4.985}{\angstrom}). The polar angle $\varphi$ defines the direction from the $\hat{\bf{x}}$ vector of the momentum \emph{k} centered around the $\overline{\rm M}$ point.
The field ${\bf\Omega}_0$ is the Rashba-like field, dominant for small momenta. The ${\bf\Omega}_1$ field has four-fold symmetry and its strength is governed by the renormalized $\lambda_{\rm nn}$ by $Q\approx (\epsilon_0 - V)^{-2}$. The ${\bf\Omega}_3$ field has a Dresselhaus-like pattern with amplitude controlled by the $Q$ renormalized product of nearest neighbor and next-nearest neighbor spin-flip hoppings. The modulus of the ${\bf\Omega}_3$ is constant in momentum space but its  spin pattern has a winding number of three. Considering the DFT calculated spin expectation values at momentum \SI{0.03}{\per\angstrom} and \SI{0.066}{\per\angstrom}, and corresponding energy splitting of \SI{0.083}{\electronvolt} and \SI{0.113}{\electronvolt} for the spin split bands, we found $\lambda_{\rm nnn}=\SI{0.078}{\electronvolt}$, $\Lambda_1=\SI{-5.77}{\electronvolt}$, and $\Lambda_3 = \SI{1.44}{\electronvolt}$. In our case  ($V\gg\epsilon_0$ and $V \approx \SI{1}{\electronvolt}$), $\lambda_{\rm nn} \approx 3\Lambda_1 / 2$, which of about 2 orders of magnitude larger than $\lambda_{\rm nnn}$, supporting spin-orbitally well bound Bi bilayers. We note that the obtained parameters were used to calculate the SOC fields simultaneously for both the momenta amplitudes, shown in Figs.~\ref{fig:Fig5}(c) and \ref{fig:Fig5}(d).

\nocite{*}%

%

\end{document}


\begin{center}
\huge{Supplemental Material for\\ \textit{A topological material in the III-V family: heteroepitaxial InBi on InAs}}\\
\normalsize
\vspace*{0.5\baselineskip}
Laurent~Nicola\"i$ ^{1,*}$, J\'an~Min\'ar$ ^{1,*}$, Maria~Christine~Richter$ ^{2,3}$, Olivier~Heckmann$ ^{2,3}$, Jean-Michel~Mariot$ ^{4}$, Uros~Djukic$ ^{2}$, Johan~Adell$ ^{5}$, Mats~Leandersson$ ^{5}$, Janusz~Sadowski$ ^{6}$, J\"urgen~Braun$ ^7$, Hubert~Ebert$ ^7$, Jonathan~D.~Denlinger$ ^{8}$, Ivana~Vobornik$ ^9$, Jun~Fujii$ ^9$, Pavol~\v{S}utta$ ^{1}$, Gavin~Bell$ ^{10}$, Martin~Gmitra$ ^{11,12}$ and Karol~Hricovini$ ^{2,3,*}$\\\
\vspace*{0.5\baselineskip}

\textit{$ ^1$ New Technologies Research Centre, University of West Bohemia, Univerzitni 8, 306 14 Pilsen, Czech Republic\\
	$ ^2$ Laboratoire de Physique des Matériaux et des Surfaces, CY Cergy-Paris Université, LIDYL, CEA, CNRS, 91191 Gif-sur-Yvette, France\\
	$ ^3$ LIDYL, Université Paris-Saclay, CEA, CNRS, 91191 Gif-sur-Yvette, France\\
	$ ^4$ Laboratoire de Chimie Physique–Matière et Rayonnement, Sorbonne Université, CNRS, 4 place Jussieu, 75252 Paris Cedex 05, France\\
	$ ^5$ MAX IV Laboratory, Lund University, Fotongatan 2, 224 84 Lund, Sweden\\
	$ ^6$ Institute of Physics, Polish Academy of Sciences, al. Lotników 32/46, 02-668 Warsaw, Poland\\
	$ ^7$ Department Chemie, Ludwig-Maximilians-Universität München, Butenandtstraße 11, 81377 München, Germany\\
	$ ^8$ Advanced Light Source, Lawrence Berkeley National Laboratory, 1 Cyclotron Road, Berkeley, CA 94720-8229, USA\\
	$ ^9$ TASC Laboratory, Istituto Officina dei Materiali, CNR, Area Science Park–Basovizza, Strada Statale 14, km 163.5, 34149 Trieste, Italy\\
	$ ^{10}$ Department of Physics, University of Warwick, CV4 7AL Conventry, United Kingdom\\
	$ ^{11}$ Institute of Physics, Pavol Jozef Šafárik University in Košice, Park Angelinum 9, 040 01 Košice, Slovak Republic\\
	$ ^{12}$ Institute of Experimental Physics, Slovak Academy of Sciences, Watsonova 47, 040 01 Košice, Slovak Republic\\}
\end{center}
\vspace*{0.5\baselineskip}

\textbf{* Corresponding authors:} lnicolai@ntc.zcu.cz, jminar@ntc.zcu.cz, karol.hricovini@cgy.fr


\vspace*{1\baselineskip}
\clearpage

\section{Low-Energy Electron Diffraction}

\renewcommand{\figurename}{}

\begin{figure}[htb]
	\centering
	\includegraphics[scale=1.1]{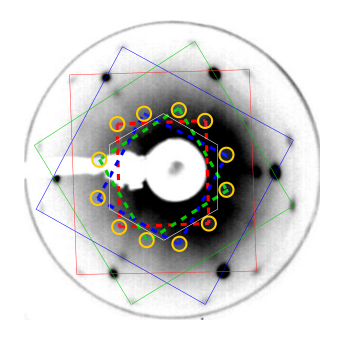}
	\captionsetup{labelformat=empty}
	\caption{Figure S1: LEED pattern: Sketch of all 3 InBi tetragonal (blue, red, green dashed lines) and Bi hexagonal (white line) BZs, with all 12 ring states (yellow circles) sitting upon the $\overline{\rm M}$ points of the InBi domains.}
	\label{fig:fig0}
\end{figure}


Figure S\ref{fig:fig0} shows the sketch of the different patterns of all 3 InBi tetragonal domains (blue, red and green). The dashed lines indicate the corresponding BZs. We additionally represent the hexagonal Bi BZ (white line) along with the ring states (yellow circles) as observed in the experimental ARPES data.

\section{InBi step heights from STM}

\begin{figure}[htb]
	\centering
	\includegraphics[scale=0.25]{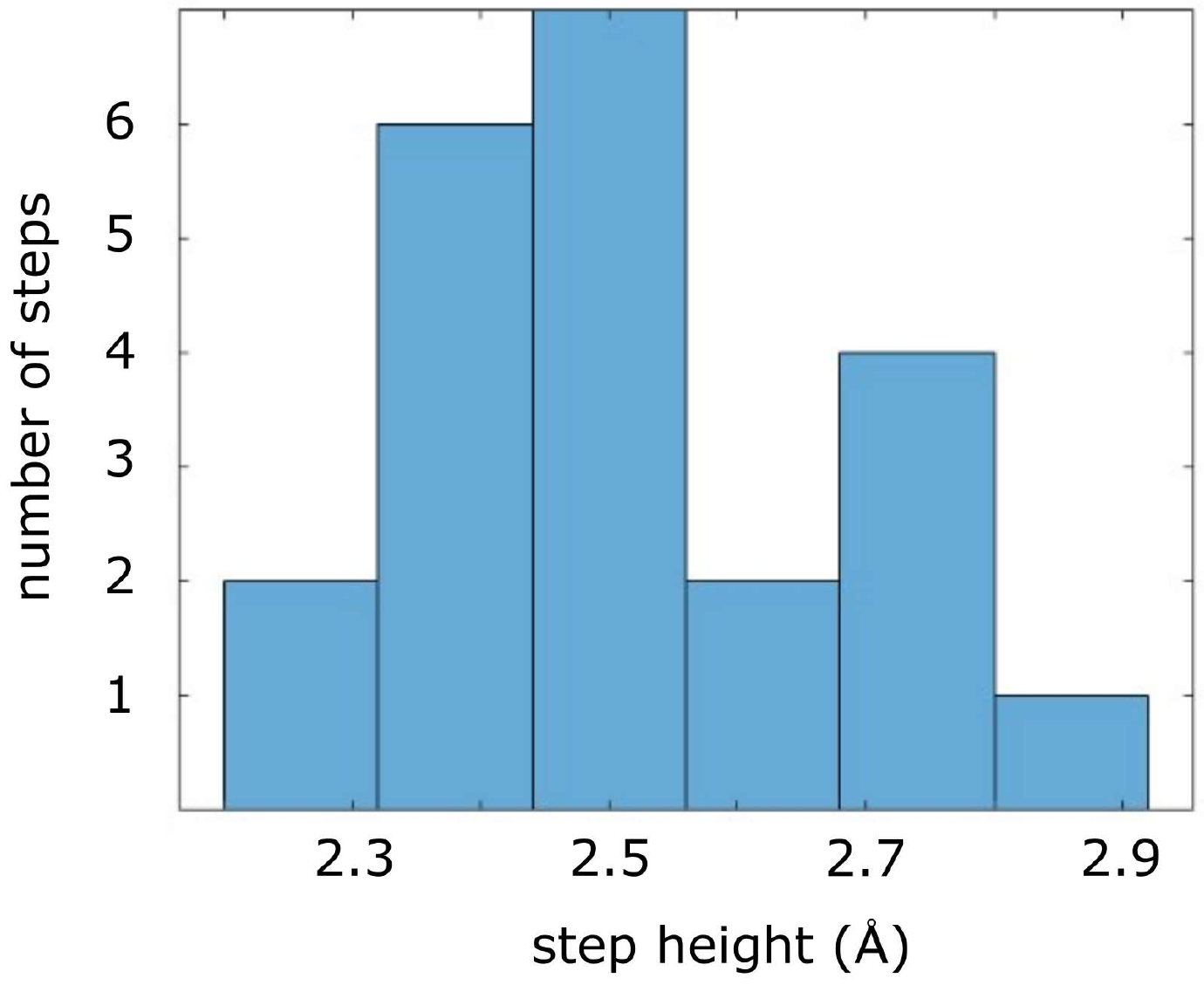}
	\captionsetup{labelformat=empty}
	\caption{Figure S2: Step height histogram for InBi(001) regions from STM. }
	\label{fig:fig0.3}
\end{figure}

The InBi regions are typically highly stepped, with terrace widths as low as a few nm and frequent step bunches. Measured step heights are affected by the levelling of such images. Nonetheless, the InBi step heights are clearly quantised by half-unit-cell height, not a full unit cell height. This is demonstrated by the histogram of measured step heights shown in Figure S\ref{fig:fig0.3}. The mean step height is 2.5 Å.

\section{Photon-dependent Photoemission Spectroscopy}

\begin{figure}[htb]
	\centering
	\includegraphics[scale=0.45]{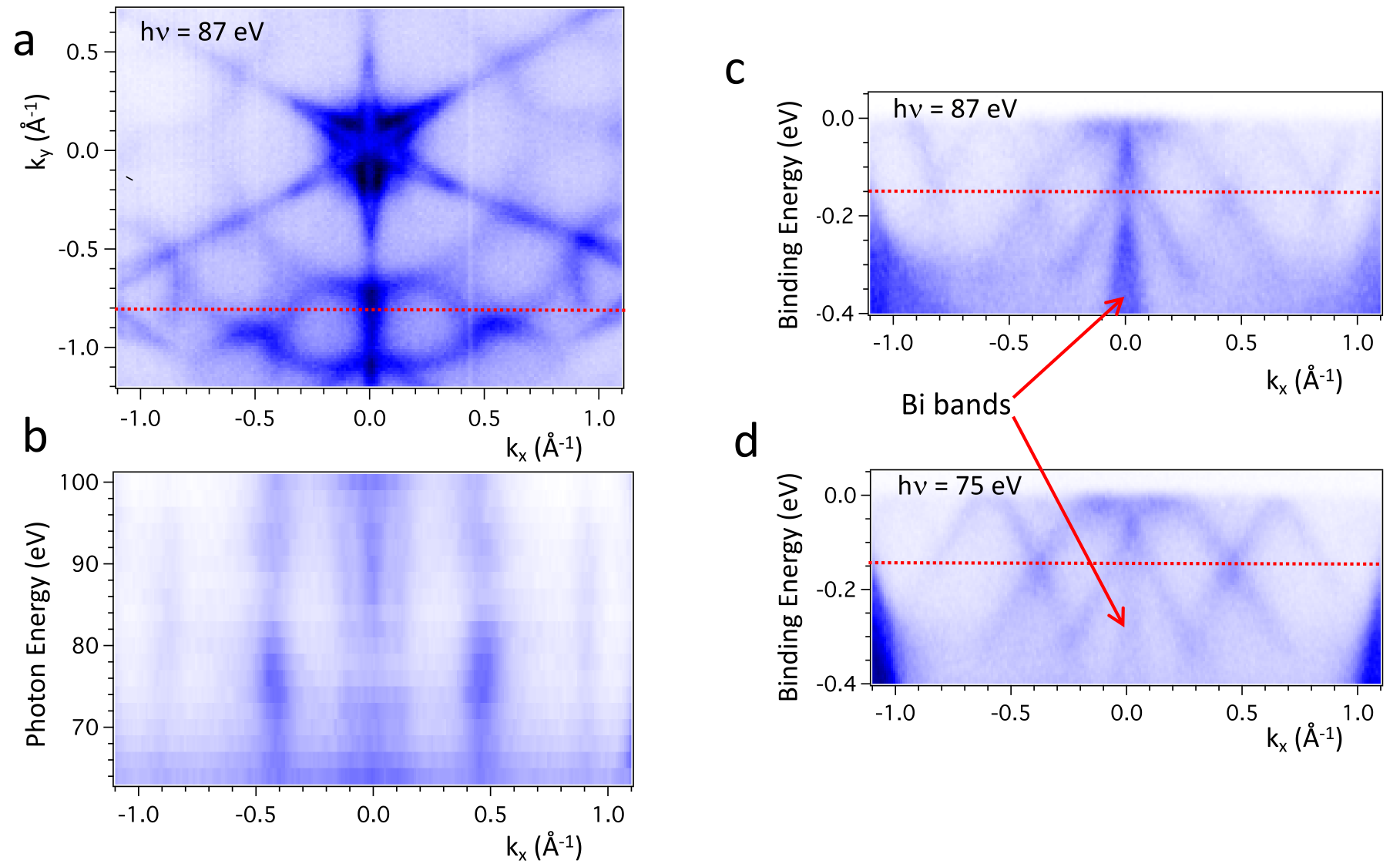}
	\captionsetup{labelformat=empty}
	\caption{Figure S3: Angle-resolved photoemission spectroscopy. Constant energy mapping of the Brillouin zone (a) and photon-energy scan (b) performed at $BE=150~meV$ as shown by red dotted lines in (c,d). (c,d) display the band dispersions at $k_{y} \approx 0.8$ \AA$^{-1}$ as indicated by red dotted line in (a). }
	\label{fig:fig0.5}
\end{figure}

Figure S\ref{fig:fig0.5} establishes the surface nature of the "ring states". Figure S\ref{fig:fig0.5}b shows that, while they are not dispersing along the $k_{z}$ direction, their intensities vary depending on the photon-energy used. It was found that, within the $[60-100]$ eV range, these states display the sharpest intensity at $h\nu=87$ eV. The very steep bands in the vicinity of $k_{x} = 0$ \AA$^{-1}$ correspond to Bi states (pointed by red arrows) in Figure S\ref{fig:fig0.5}c,d. Therefore, in Figure S\ref{fig:fig0.5}b, the line at $k_{x} = 0$ \AA$^{-1}$ correspond to Bi while the 6 other lines correspond to InBi topological surface states.

\section{Photoemission signal from various sample locations}

\begin{figure}[htb]
	\centering
	\includegraphics[scale=0.5]{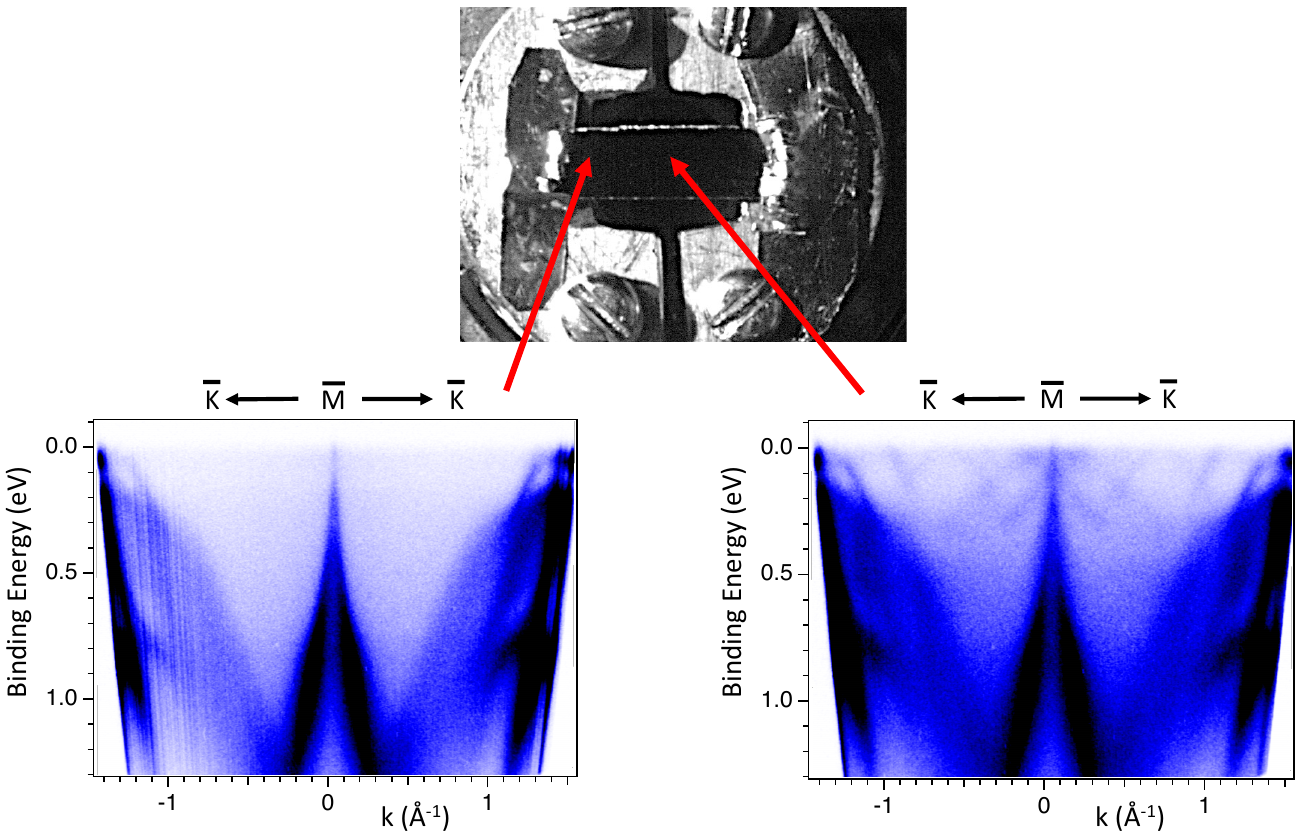}
	\captionsetup{labelformat=empty}
	\caption{Figure S4: Configuration of the sample mount (top). ARPES spectra  (bottom) measured at two different locations, as shown by arrows.}
	\label{fig:fig0.8}
\end{figure}

The InAs(111)-A substrate was annealed by direct current heating. For this purpose metal clamps were insuring the current injection into the sample as depicted in Figure S\ref{fig:fig0.8} (top). This configuration can lead to a non-uniform distribution of the electric current (and so of the temperature), especially close to the metal clamps on both ends of the sample, due to randomly established electric contacts. Thus, the In atoms reservoir through the surface is as well not completely homogeneous. In the example shown, approximately 20 bilayers of Bi were deposited, with a thickness considered as constant over the entire sample as the evaporation cone was much larger than the sample dimensions. Therefore, if present, the strain effects on the electronic structure can reasonably be considered as constant throughout the sample surface. After evaporation of Bi, we observed InBi domains only towards the centre of the sample, whereas only Bi domains were present on the sides. This is illustrated by the corresponding spectra in Figure S\ref{fig:fig0.8} bottom, showing the measured electronic structure along the $\overline{\rm K}$--$\overline{\rm M}$--$\overline{\rm K}$ direction [referred to the Bi(111) surface]. Only the bottom right spectrum shows the ring states (i.e. the topological surface states) coming from the InBi domains, as shown in Figure 3 (d) and (e) of the main manuscript. As seen, the Bi bands exhibit the same shape in both spectra.

\begin{figure}[htb]
	\centering
	\includegraphics[scale=0.45]{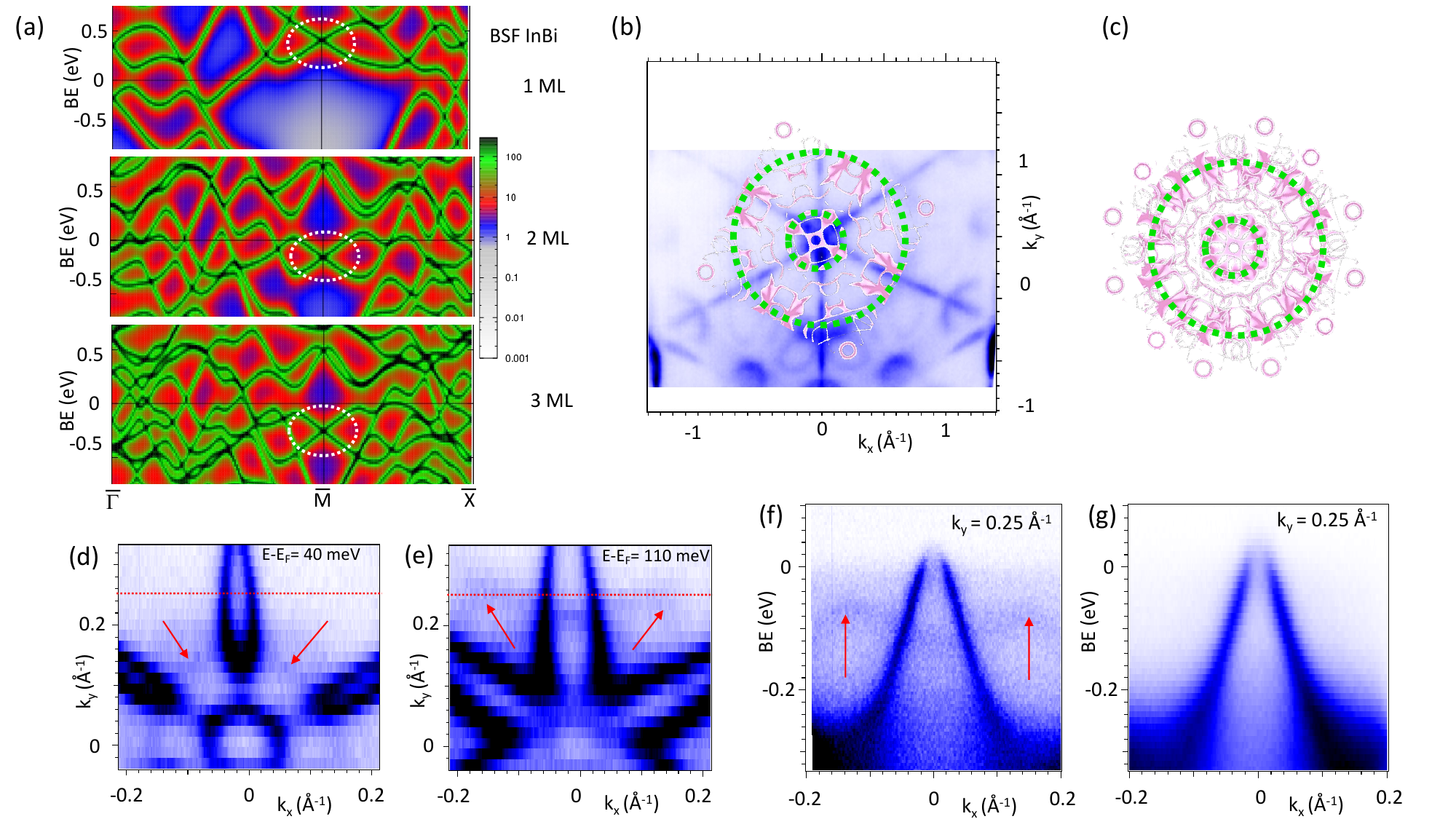}
	\captionsetup{labelformat=empty}
	\caption{Figure S5: Electronic band structure from Bi and InBi domains. (a) Bloch Spectral Function depicting the electronic structure along the $\overline{\rm \Gamma}$--$\overline{\rm M}$--$\overline{\rm X}$ directions for 1, 2 and 3 InBi free-standing layers. (b) Measured Fermi surface, overlaid with the iso-energetic BSF of the InBi 3 layers case (pink). Two green dotted circles are guides to the eye (see the text for discussion). (c) Sketch of the electronic structure with 3 InBi domains (pink) with the resulting circular structures (marked by green dotted lines). Iso-energetic spectra on the Bi+InBi domains at (d) $-40~meV$ and (e) $-110~meV$.  (f) ARPES spectrum along the $\overline{\rm \Gamma}$--$\overline{\rm K}$ direction [see red dotted lines in (d) and (e)]. The red arrows point InBi bands. (g) ARPES spectrum of pure Bi crystal.}
	\label{fig:fig0.9}
\end{figure}

The InBi topological states close to the $\overline{\rm M}$ points are very robust and are easily detectable in the corresponding band gap (see Figure S\ref{fig:fig0.9}). Ground state Bloch spectral function (BSF) calculations show their presence (indicated by white dotted loops) even in free-standing thin layers, as shown in Figure S\ref{fig:fig0.9} (a) for 1, 2 and 3 InBi monolayers. Outside the $\overline{\rm M}$ point band gap, the InBi crystal is metallic with many bands appearing at the Fermi surface.
The presence of domains in three epitaxial orientations is mixing the signal from the InBi crystals in the ARPES measurements. In Figure S\ref{fig:fig0.9} (b) we overlap the constant energy map [presented in Figure 3 (a) of the manuscript] with the calculated one domain Fermi surface for 3 ML. Clearly, the positions of the topological states fit well with the measurements. When generating the 3-domain situation [Figure S\ref{fig:fig0.9} (c)] InBi metallic bands collapse in a complex array forming roughly 2 rings (indicated by green circles). At the same time, the topological states are nicely separated and match with the experimental observation. 
The outer ring of InBi bands is detected in the ARPES spectrum; the presence of the inner band ring is proven in Figures S\ref{fig:fig0.9} (d) and (e). Constant energy maps at $-40~meV$ and $-110~meV$ below the Fermi level and close to the $\rm \Gamma$ point reveal a set of bands (red arrows) that are not belonging to the Bi(111) monocrystal. This is further evidenced in Figure S\ref{fig:fig0.9} (f) where the ARPES E(k) dispersion at ky = 0.25 Å-1 [red dotted horizontal lines in Figure S\ref{fig:fig0.9} (d) and (e)] is shown. A new band (red arrows), belonging to InBi, is crossing the strong intensity dispersion of the Bi hole pocket. Figure S\ref{fig:fig0.9} (g) shows a spectrum of a pure Bi(111) monocrystal taken at the same $k$-point where this band is clearly missing.

\break

\section{Bloch Spectral Function: $k_z$ scan}

\begin{figure}[htb]
	\centering
	\includegraphics[scale=0.9]{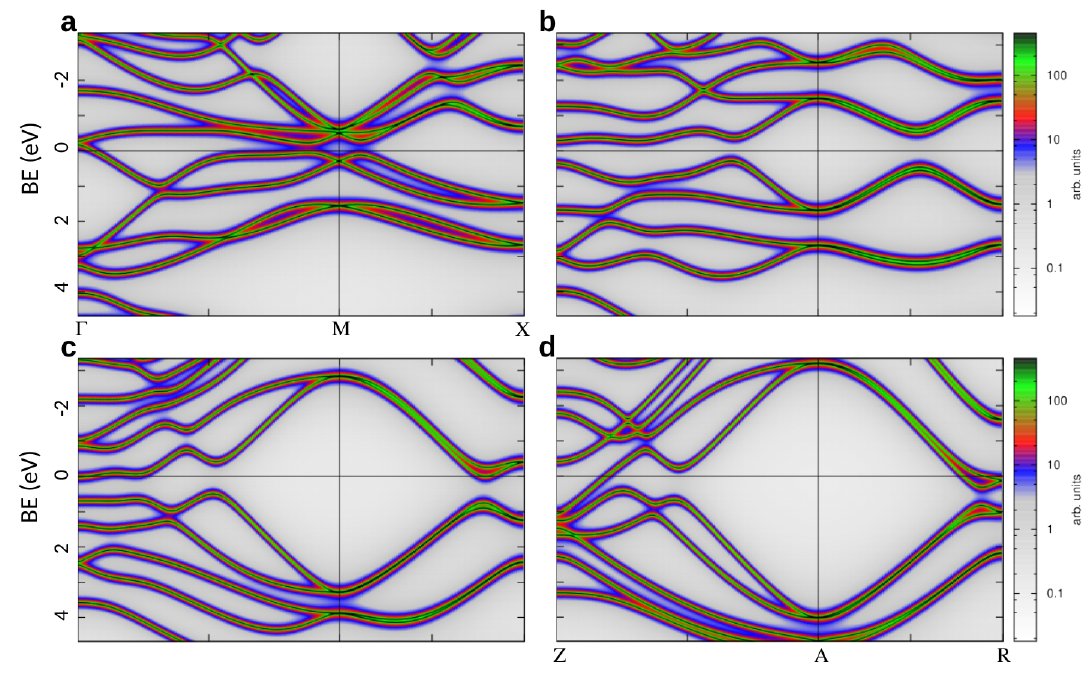}
	\captionsetup{labelformat=empty}
	\caption{Figure S6: Bloch Spectral Function. Bulk band dispersion in the 3D BZ (a-d) going from the $\rm \Gamma M X$ plane to the $\rm Z A R$ plane with a constant step size. }
	\label{fig:fig1}
\end{figure}

Figure S\ref{fig:fig1} shows a $k_z$ scan (along the [001] direction). The calculations are done for an infinite crystal and contain therefore only bulk states. At the $\text{M}$ point, there is a bulk band gap smaller than 1eV close to the Fermi level. When going towards the A point, the gap not only survives but widens up to almost 7 eV.   

\section{Atomic contributions within InBi}

\begin{figure}[htb]
	\centering
	\includegraphics[scale=0.9]{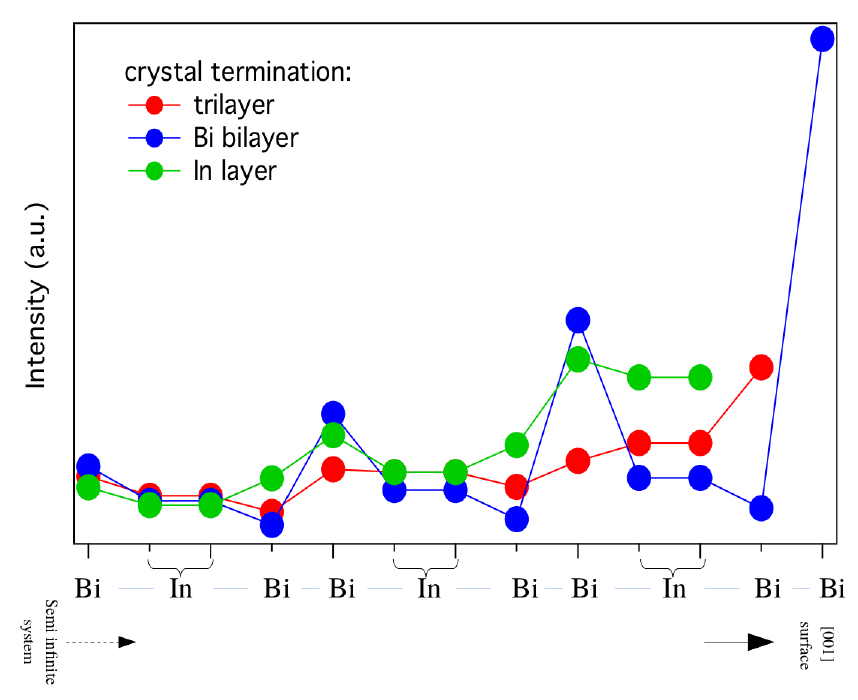}
	\captionsetup{labelformat=empty}
	\caption{Figure S7: Atomic contributions. Layer-resolved contributions for all 3 possible surface terminations: Bi-bilayer (blue), Bi-monolayer (red) and In (green).}
	\label{fig:fig2}
\end{figure}

The contribution of the different atoms in the intensity at the Dirac-like point at $\overline{\text{M}}$ in shown in Figure S\ref{fig:fig2}. In all cases, the Bi layer located the closest to the surface is contributing the most to the crossing point. We also noticed that two consecutive Bi always depict an asymmetric contribution. This behaviour suggests a coupling of the 2 Bi atoms. Note that two atoms of In are here represented because the In atomic plane is composed of two non-equivalent (two Wyckoff) positions. Their contributions are identical in the vicinity of the $\overline{\rm M}$ point.

\newpage




\newpage

\clearpage